\documentclass{Interspeech}

\interspeechcameraready

\title{Bayesian Learning for Domain-Invariant Speaker Verification and Anti-Spoofing}

\author[affiliation={1,2}]{Jin}{Li}
\author[affiliation={1}]{Man-Wai}{Mak}
\author[affiliation={2}]{Johan}{Rohdin}
\author[affiliation={1}]{Kong Aik}{Lee}
\author[affiliation={2}]{Hynek}{Hermansky}

\affiliation{Department of Electrical and Electronic Engineering}{The Hong Kong Polytechnic University}{Hong Kong SAR}
\affiliation{Speech@FIT}{Brno University of Technology}{Czechia}
\email{jin666.li@connect.polyu.hk}

\usepackage{comment}

\usepackage{booktabs,bm,pifont,multirow}
\usepackage{bbding}
\usepackage{pifont}
\usepackage{array,booktabs}
\usepackage{bm}
\usepackage{algorithm}
\usepackage{array}
\usepackage{stfloats}
\usepackage{url}
\usepackage{verbatim}
\usepackage{mathtools}
\usepackage{booktabs}
\usepackage{tablefootnote}
\usepackage{lscape}
\usepackage{cite}
\usepackage{fullwidth}
\usepackage{verbatim}
\usepackage{cite}
\usepackage{amsmath,amssymb,amsfonts}
\usepackage{algorithmic}
\usepackage{graphicx}
\usepackage{textcomp}
\usepackage{xcolor}
\usepackage{arydshln} 
\usepackage{amsthm}

\begin{document}
	
	\maketitle
	
	\begin{abstract}
		
		The performance of automatic speaker verification (ASV) and anti-spoofing drops seriously under real-world domain mismatch conditions. The relaxed instance frequency-wise normalization (RFN), which normalizes the frequency components based on the feature statistics along the time and channel axes, is a promising approach to reducing the domain dependence in the feature maps of a speaker embedding network. We advocate that the different frequencies should receive different weights and that the weights' uncertainty due to domain shift should be accounted for. To these ends, we propose leveraging variational inference to model the posterior distribution of the weights, which results in Bayesian weighted RFN (BWRFN). This approach overcomes the limitations of fixed-weight RFN, making it more effective under domain mismatch conditions. Extensive experiments on cross-dataset ASV, cross-TTS anti-spoofing, and spoofing-robust ASV show that BWRFN is significantly better than WRFN and RFN.
	\end{abstract}

	\section{Introduction}
	Automatic speaker verification (ASV) seeks to authenticate a speaker’s identity by analyzing their voice \cite{mak2020machine,li2023modeling}. It is widely applied across various applications, including biometric authentication on personal smart devices \cite{lee2013}. Recently, inspired by the robust feature extraction abilities of deep neural networks (DNNs), numerous deep learning-based speaker recognition methods have been introduced \cite{bai2021speaker}.
	
	However, ASV systems deployed in real-world applications are negatively affected by domain mismatch between the training and test conditions due to various factors, such as variations in recording devices, acoustic environment, audio channels, and languages. In addition, ASV systems are vulnerable to unseen spoofing attacks, which are often created by new algorithms that are not seen by the systems during the training phase \cite{zeng2024spoofing}. Domain adaptation (DA) \cite{9134370} and domain generalization (DG) \cite{zhou2022domain} are two mainstream approaches to address the domain mismatch problem. 
	
	DA has been developed to address domain shift problems by transferring knowledge from a well-labeled, resource-rich source domain to a target domain \cite{Lee2019,wang2018unsupervised,garcia2014unsupervised}. For example, a domain adversarial training approach addresses the domain mismatch problem by projecting various domains into the same subspace \cite{wang2018unsupervised}, thereby becoming domain invariant across seen domains. A probabilistic linear discriminant analysis (PLDA) adaptation was used to cluster unlabeled in-domain data and then use this data to adapt the parameters of out-of-domain models \cite{garcia2014unsupervised}. Correlation alignment (CORAL) was another widely used backend method that aligns out-of-domain statistics with those of the in-domain statistics \cite{Lee2019}.
	
	On the other hand, domain generalization learns a model that remains robust to domain shift without requiring adaptation to the target domains \cite{li2018learning}. Numerous approaches have been proposed to address the domain generalization issue by achieving strong generalization to unseen test domains in recent years. For example, the meta-generalized speaker verification was introduced via meta-learning to improve the generalization ability of the model on unseen domains \cite{zhang2023meta}. In \cite{li2023mutual}, a mutual information-based embedding decoupling method was proposed to improve domain generalization capabilities. More recently, Relaxing Instance Frequency-wise Normalization (RFN) was incorporated explicitly into the model to eliminate instance-specific domain discrepancies in acoustic features while minimizing the loss of useful discriminative information \cite{kim2022domain}.
	
	However, previous works are still prone to overfitting and poor generalization because the use of fixed parameters during inference fails to account for model uncertainty, leading to overfitting on the source domains \cite{abdar2021review}. Although it was shown in \cite{kim2022domain} that frequency-wise distribution is highly correlated with domain information, the approach suffers from overfitting on the source domain because the uncertainty of weights for layer normalization and instance frequency-wise normalization is not taken into account.
	
	In this paper, we address domain generalization by extending prior work in \cite{kim2022domain} under a probabilistic framework. To better explore domain-invariant learning, we model the uncertainty in the weighting parameters of RFN by leveraging variational Bayesian inference. Bayesian learning has been applied to solve many machine learning problems in speech community \cite{lam2018gaussian,xue2022bayesian,DBLP:conf/odyssey/LiZYHWLM20}. It was applied to model the uncertainty on different normalizers to cope with the model's uncertainty. To better explore domain-invariant learning, we introduce uncertainty to the weights of the normalized frequency components in the convolutional feature maps of a speaker embedding network by leveraging variational Bayesian inference. This enables us to explore domain invariance in a principled way to achieve domain-invariant feature representations and speaker classification jointly. In our experiments, we directly compare the performance of a Bayesian learning-based neural network with a non-Bayesian learning-based neural network. Additionally, we evaluate our methods against various baselines in multiple tasks, including ASV, anti-spoofing, and SASV.
	
	\section{Preliminaries}
	In this section, we define the notations and present the preliminary on \textit{relaxed instance frequency-wise normalization}.
	
	\subsection{Relaxed Instance Frequency-wise Normalization}
	An audio signal processing system typically applies spectral-domain analysis on each audio channel on a frame-by-frame basis. Previous studies on audio features \cite{kim2022domain} have shown that the frequency features (vectors whose elements are the frequency components) contain more domain-relevant information than the vectors defined by the channel axis. Therefore, domain mismatch could be reduced by reducing the domain-dependent variability along the frequency axis. One approach to reducing this variability is to normalize the frequency components in a channel-independent manner using the statistics obtained from the entire audio signal. The authors in \cite{kim2022domain} named the method ``Instance Frequency-wise Normalization (INF).'' 
	
	Given a mini-batch of $N$ multichannel audio signals, its audio characteristics in the frequency domain can be represented by a tensor $\bm{x}\in\mathbb{R}^{N\times C\times F\times T}$, where $N$, $C$, $F$, and $T$ are the mini-batch size, numbers of channels, frequency bins, and frames, respectively. The IFN of $\bm{x}$ at channel $c$ and frame $t$ is defined as:
	\begin{equation}
		\label{eq:4}
		\mbox{IFN}(\boldsymbol{x})_{:,c,:,t} = \frac{\boldsymbol{x}_{:,c,:,t} - \mathbb{E}_{\scriptscriptstyle{\text{IFN}}}(\boldsymbol{x})}{\sqrt{\mathbb{V}_{\scriptscriptstyle{\text{IFN}}}(\boldsymbol{x})+\epsilon}} \in\mathbb{R}^{N\times F} \, ,
	\end{equation}
	where $\mathbb{E}_{\scriptscriptstyle{\text{IFN}}}$ and $\mathbb{V}_{\scriptscriptstyle{\text{IFN}}}$ are the first and second order statistics given by 
	\begin{equation}
		\label{eq:4a}
		\mathbb{E}_{\scriptscriptstyle{\text{IFN}}}(\boldsymbol{x}) = \frac{1}{C\cdot T} \sum\nolimits_c^C\sum\nolimits_t^T \boldsymbol{x}_{:,c,:,t} \, 
	\end{equation}
	and
	\begin{equation}
		\label{eq:4b}
		\begin{split}
			\mathbb{V}_{\scriptscriptstyle{\text{IFN}}}(\boldsymbol{x}) = \frac{1}{C\cdot T} \sum\nolimits_c^C\sum\nolimits_t^T &(\boldsymbol{x}_{:,c,:,t} - \mathbb{E}_{\scriptscriptstyle{\text{IFN}}}(\boldsymbol{x})) \\  &\odot (\boldsymbol{x}_{:,c,:,t}-\mathbb{E}_{\scriptscriptstyle{\text{IFN}}}(\boldsymbol{x})) \, ,
		\end{split}
	\end{equation}
	respectively. Here, $\epsilon$ is a small positive constant, and $\odot$ represents the Hadamard product. Define the layer normalization (LN) at channel $c$, frequency $f$, and frame $t$ as:
	\begin{equation}
		\label{eq:5}
		\mbox{LN}(\boldsymbol{x})_{:,c,f,t} = \frac{\boldsymbol{x}_{:,c,f,t} - \mathbb{E}_{\scriptscriptstyle{\text{LN}}}(\boldsymbol{x})}{\sqrt{\mathbb{V}_{\scriptscriptstyle{\text{LN}}}(\boldsymbol{x})+\epsilon}} \, \in\mathbb{R}^N \, ,
	\end{equation}
	where $\mathbb{E}_{\scriptscriptstyle{\text{LN}}}(\boldsymbol{x})=\frac{1}{C\cdot F\cdot T} \sum_{c}^{C} \sum_{f}^{F} \sum_{t}^{T} \boldsymbol{x}_{:,c,f,t}$, and $\mathbb{V}_{\scriptscriptstyle{\text{LN}}}(\boldsymbol{x})=\frac{1}{C\cdot F\cdot T} \sum_{c}^{C} \sum_{f}^{F} \sum_{t}^{T} (\boldsymbol{x}_{:,c,f,t}-\mathbb{E}_{\scriptscriptstyle{\text{LN}}}(\boldsymbol{x}))\odot (\boldsymbol{x}_{:,c,f,t}-\mathbb{E}_{\scriptscriptstyle{\text{LN}}}(\boldsymbol{x}))$. Then, the Relaxed instance Frequency-wise Normalization (RFN) is calculated as follows:
	\begin{equation}
		\label{eq:6}
		F(\boldsymbol{x}) = \lambda \mbox{LN}(\boldsymbol{x}) + (1-\lambda) \mbox{IFN}(\boldsymbol{x}) \in\mathbb{R}^{N\times C\times F\times T} \, ,
	\end{equation}
	where the scalar $\lambda \in [0,1]$ denotes the degree of relaxation.
	
	\section{Bayesian Relaxed Instance Frequency-wise Normalization}
	Eq.~\ref{eq:6} interpolates between IFN and LN style normalization. However, the degree of relaxation, $\lambda$, of individual frequency bins are fixed and equal. We hypothesize that different frequency bins have different levels of domain dependence and should be weighted differently. We therefore extend RFN to weighted RFN (or WRFN), as follows
	\begin{equation}
		\label{eq:8b}
		\begin{split}
			F(\boldsymbol{x})=& \lambda \mbox{LN}(\boldsymbol{x})\odot \sigma(\boldsymbol{w}_1)  + (1-\lambda) \mbox{IFN}(\boldsymbol{x})\odot \sigma(\boldsymbol{w}_2) \\
			&\in \mathbb{R}^{N\times C\times F\times T} \, ,
		\end{split}
	\end{equation}
	where $\boldsymbol{w}_1,\boldsymbol{w}_2 \in \mathbb{R}^{F}$ are broadcast along the mini-batch, time and channel axes, $\sigma(\cdot)$ is a sigmoid function that squashes the weights to values between 0 and 1, WLN is weighted LN, and WIFN is weighted IFN. We define $\boldsymbol{w}=[\boldsymbol{w}_1^\mathsf{T},\boldsymbol{w}_2^\mathsf{T}]^\mathsf{T}$, where $\mathsf{T}$ is the transpose. Note that the frequency specific weights, $\lambda\sigma(\boldsymbol{w}_1)$ and $\lambda\sigma(\boldsymbol{w}_2)$ generally does not sum to one. However, this can to a large extent be compensated by the subsequent convolutionary layer.
	
	The weights in WRFN, $\boldsymbol{w}_1$ and $\boldsymbol{w}_2$, are deterministic and learned from training data. Since the purpose of these weights is to address domain variations and since there number of domains in the training data is scarce, we hypothesize that the weights will be prone to overfitting. To mitigate this, we improve the robustness of the weighted RFN (Eq.~\ref{eq:8b}) by modeling the uncertainty in ${\boldsymbol{w}}$ with Bayesian techniques, which results in Bayesian Weighted Relaxed Instance Frequency-wise Normalization (BWRFN).	
	
	To account for uncertainty in $\boldsymbol{w}$, we assume that the weight vector is drawn from a prior distribution $p(\boldsymbol{w})$. Let ${\cal D}=\{(\bm{x}_i,\bm{s}_i)\}_{i=1}^N$ denote the \emph{training set}, considering of $N$ utterances, each spoken by one of the $S$ speakers. As defined in Eqs.~\ref{eq:4}--\ref{eq:8b}, each tensor $\bm{x}_i\in\mathbb{R}^{C\times F\times T}$ contains the acoustic features (e.g., filterbank features) of an utterance, and the corresponding speaker ID is represented as a one-hot vector $\bm{s}_i\in\mathbb{R}^S$. To predict the speaker label (or generate a speaker embedding) for a \emph{test utterance}, we require the posterior distribution $p(\boldsymbol{w}|\mathcal{D})$ along with the other model parameters. Since computing the posterior $p(\bm{w}|{\cal D})$ is intractable, we employ variational inference (see e.g. Ch. 10 of \cite{bishop2006pattern}) to approximate it. Specifically, $p(\bm{w}|{\cal D})$ is approximated by a Gaussian distribution $q(\bm{w})$ with mean $\boldsymbol{\mu}$ and diagonal covariance $\boldsymbol{\sigma}^2$. The KL divergence, $D_{KL}(q(\boldsymbol{w})||p(\bm{w}|{\cal D}))$, can be minimized jointly with maximization of the (conditional) marginal log-likelihood of the training data, $p\left(\{\boldsymbol{s}_i\}_{i=1}^N|\{\boldsymbol{x}_i\}_{i=1}^N \right)$, by maximizing the variational lower bound $\mathcal{L}(\boldsymbol{\theta}, \boldsymbol{\mu}, \boldsymbol{\sigma})$ as described in \cite{kingma2015variational}:	
	\begin{align}
		\label{eq:11}
		\mathcal{L}(\boldsymbol{\theta},\boldsymbol{\mu},\boldsymbol{\sigma})
		=&\overbrace{\sum_{i=1}^{N} \mathbb{E}_{\boldsymbol{w}\sim  q(\boldsymbol{w})}  \log p_{\boldsymbol{\theta}}(\boldsymbol{s}_i|\boldsymbol{x}_i,\boldsymbol{w})}^{\substack{\mathcal{L}_1}} \nonumber\\
		&\,\,\,\,- D_{KL}(q(\boldsymbol{w})||p(\boldsymbol{w})).
	\end{align} 
	where $p_{\boldsymbol{\theta}}(\boldsymbol{s}_i|\boldsymbol{x}_i,\boldsymbol{w})$ is the speaker probabilities given by softmax layer of the neural network and $\bm \theta$ are the parameters of the neural network (other than $\bm w$). The prior, $p(\bm{w})$ is assumed to be a standard Gaussian, i.e., $p(\bm{w})={\cal N}(\bm{w}; \bm{0},\bm{I})$. On the other hand, the variational posterior $q(\bm{w})$ is modeled as a diagonal Gaussian distribution, i.e., $q(\bm{w})={\cal N}(\bm{w};\bm{\mu}_{\bm{w}},\bm{\Sigma}_{\bm{w}})$, where the covariance matrix is defined as $\bm{\Sigma}_{\bm{w}}=\text{diag}(\boldsymbol{\sigma}^2_{\boldsymbol{w}})$. The diagonal Gaussian distribution is initialized randomly but learned during training to approximate the true posterior $p(\bm{w}|{\cal D})$. The first term of Eq.~\ref{eq:11} is the expectation of log-likelihood of the data over the approximated posterior distribution $q(\boldsymbol{w})$. This expectation can be computed by using Monte Carlo sampling, i.e.,
	\begin{equation}
		\label{eq:17}
		\begin{split}
			{\cal L}_1 &=\sum_{i=1}^N\int_{\boldsymbol{w}} \log p_{\boldsymbol{\theta}}(\boldsymbol{s}_i|\boldsymbol{x}_i,\boldsymbol{w})q(\boldsymbol{w})\, \mathrm{d} \boldsymbol{w} \nonumber \\
			&\approx\frac{1}{K}\sum_{k=1}^K\sum_{i=1}^N\left[\log p_{\boldsymbol{\theta}}(\bm{s}_i|\bm{x}_i,\bm{w}^{(k)})\right] \, ,
		\end{split}
	\end{equation}
	where $K$ denotes the number of samples and $\boldsymbol{w}^{(k)}$ is the $k$-th sample drawn from the distribution $q(\boldsymbol{w})$. 
	
	For sampling $\boldsymbol{w}^{(k)}$, the re-parametrization trick \cite{kingma2015variational} is applied to produce the $i$-th element $w_i^{(k)}$ of $\boldsymbol{w}^{(k)}$ as follows:
	\begin{equation}
		w_i^{(k)}=\mu_{\bm{w},i}+\sigma_{\bm{w},i}\epsilon^{(k)}_i,~~\epsilon^{(k)}_i\sim{\cal N}(0,1) \, ,
	\end{equation}
	where $\mu_{\bm{w},i}$ is the $i$-th element of $\bm{\mu}_{\bm{w}}$ and $\sigma_{\bm{w},i}$ is the square root of the $i$-th diagonal element of $\bm{\Sigma}_{\bm{w}}$. 
	
	By assuming Gaussians for $p(\boldsymbol{w})$ and $q(\boldsymbol{w})$, the second term of Eq.~\ref{eq:11} has a closed-form solution. It can be derived as follows:
	\begin{equation}
		\label{eq:18}
		\begin{split}
			\mathcal{L}_{\text{KL}} &= -D_{KL}(q(\boldsymbol{w})||p(\boldsymbol{w})) \\
			&=-D_{KL} (q(\boldsymbol{w})|\mathcal{N}(\boldsymbol{w};\boldsymbol{0},\boldsymbol{I}))  \\
			&= \frac{1}{2} \displaystyle\sum_{f=1}^{F} \left[  1+\log (\sigma_{\bm{w},f}^2)-\sigma_{\bm{w},f}^2-\mu_{\bm{w},f}^2  \right] \, ,
		\end{split}
	\end{equation}
	where $\sigma_{\bm{w},f}$ and $\sigma_{\bm{w},f}$ are the parameters of the variational posterior, and $f$ represents index of frequency bins $F$.
	
	During inference, posterior predictive uncertainty consists of aleatoric and epistemic components. Epistemic uncertainty stems from the limited amount of domains used to estimate the model, while aleatoric uncertainty arises from the inherent randomness in the data. The posterior predictive distribution:
	\begin{equation}
		\label{eq:19}
		\begin{split}
			\hat{\bm y}&=\int_{\boldsymbol{w}} p_{\boldsymbol{\theta}^*}(\hat{y}(\boldsymbol{z})|\boldsymbol{w}) p(\boldsymbol{w}|\mathcal{D})\,\mathrm{d}\boldsymbol{w} \\
			&=\mathbb{E}_{\boldsymbol{w}\sim p(\boldsymbol{w}|\mathcal{D})} \left[ p_{\boldsymbol{\theta}^*}(\hat{y}(\boldsymbol{z})|\boldsymbol{w}) \right] \, ,
		\end{split}
	\end{equation}
	where $\hat{y}(\cdot)$ represents the embedding layer output, $\bm \theta^*$ is an optimal parameters of the neural network (other than $\bm w$), and $\boldsymbol{z}$ is the Mel-spectrogram of a new utterance. This distribution can be expressed as the expectation of the single network likelihood under the posterior $p(\boldsymbol{w}|\mathcal{D})$, interpreting the predictive distribution as an infinite ensemble of network's outputs \cite{blundell2015weight}. Each network's output contribution is weighted by the posterior of the weights given to the training data. This infinite ensemble can be approximated using a finite number of Monte Carlo samples from the posterior. For simplicity, we use the expected value of embedding layer output during inference. 
	
	\section{Experimental Setup}
	
	\subsection{Evaluation Tasks}
	The experiments were conducted on cross-dataset ASV tasks, anti-spoofing tasks, and spoofing-robust automatic speaker verification (SASV) tasks.
	
	\subsubsection{Cross-Dataset Speaker Verification}
	We followed \cite{zhang2023meta} for cross-dataset speaker verification. To assess robustness, we designed a setup using CN-Celeb \cite{li2022cn} (testing genre mismatches), HI-MIA \cite{qin2020hi} (reflecting device and environment variability), FFSVC \cite{qin2020interspeech} (containing cross-channel far-field audio), and Voxceleb \cite{nagrani2020voxceleb} with 1,307 training speakers in total. Our trials simulate cross-genre and cross-device conditions to comprehensively evaluate genre, device, and dataset mismatches. During training, HI-MIA was treated as the unseen domain, while the model was trained on the remaining datasets.
	
	\subsubsection{Cross-TTS Anti-spoofing}
	In practice, attackers may employ various TTS, voice conversions (VC), and adversarial methods to generate spoofed speech, leading to domain mismatches between the training, development, and evaluation sets. In the closed condition of ASVspoof 5 Track~1 \cite{wang24_asvspoof}, participants were limited to using only the provided training and development data. Following \cite{wang24_asvspoof}, we built our systems on the training and development set and evaluated them on the evaluation set (see Table~2 of \cite{wang24_asvspoof}).

	\subsubsection{Cross-TTS SASV}	
	ASVspoof 5 Track~2 \cite{wang24_asvspoof} evaluates spoofing-robust ASV (SASV) systems by simulating a telephony scenario where synthetic and converted speech was directly injected into a telephone line. Participants may develop either standalone classifiers or fuse separate ASV and countermeasure (CM) subsystems.
	
	\subsection{Implementation Details for Cross-dataset Speaker Verification}
	
	We used data augmentation to train speaker embedding networks, including adding music, noise, and babble from MUSAN \cite{snyder2015musan} and convolving the original speech with the room impulse responses from RIR \cite{allen1979image}. The original and augmented utterances were then cut into 2-second segments. For each segment, we extracted 40-dimensional mel-filterbank features using a 25ms window with a frameshift of 10ms. The filterbank features were then presented to different front-ends, as described below. 
	\begin{itemize}		
		\item {\it R-vector}. The R-vector network \cite{qin2019far,wang2020data} employs ResNet-18 to transform the mel-filterbank features into deep features, which were then mapped to 256-dimensional embeddings via average pooling. During training, softmax was used to compute the cross-entropy loss. The model was trained for 100 epochs using SGD with a momentum of 0.9, a weight decay of 0.0001, a mini-batch size of 100, and a learning rate decayed every 10 epochs from 0.1.
		
		\item {\it ECAPA-TDNN}. We also used ECAPA-TDNN \cite{desplanques2020ecapa} to extract speaker embeddings, using cross-entropy loss (softmax) to optimize the network.
		
		\item {\it RFN-R-vector}. We constructed an RFN-R-vector network by inserting an RFN operation (Eq.~\ref{eq:6}) before the first convolution layer and after each residual block of the R-vector network. All other training settings remain the same as those in the R-vector, with $\lambda$ in Eq.~\ref{eq:6} set to 0.5 as in \cite{kim2022domain}.	
		
		\item {\it WRFN-R-vector}. A WRFN-R-vector network is obtained by weighting the normalized frequency components (Eq.~\ref{eq:8b}) in an RFN-R-vector network for each RFN operation.
		
		\item {\it BWRFN-R-vector}. We built a BWRFN-R-vector network by integrating Bayesian learning into the frequency normalization weights (Eqs.~\ref{eq:17}--\ref{eq:18}) of the WRFN-R-vector network while keeping the same training settings as the RFN-vector network. For efficiency, a single Monte Carlo sample was used for each 4-second speech segment (i.e., $K=1$ in Eq.~\ref{eq:17}, similar to the setting in \cite{kingma2013auto}.). Additionally, we replaced the WRFN operations in the WRFN-R-vector network with BWRFN operations.
	\end{itemize}
	
	\subsection{Implementation Details for Cross-TTS Anti-spoofing}
	{\it BWRFN-ResNet} was based on ResNet18 \cite{rohdin24_asvspoof} with the addition of BWRFN, where the placement of BWRFN was consistent with that in {\it BWRFN-R-vector}. We applied extensive data augmentation during training, using MUSAN's \cite{snyder2015musan} noise subset, RIR \cite{allen1979image}, RawBoost \cite{tak2022rawboost}, Audiomentations,\footnote{\href{https://github.com/iver56/audiomentations}{https://github.com/iver56/audiomentations}} and codecs \cite{okhotnikov2024idvoice}. 
	
	\subsection{Implementation Details for Cross-TTS SASV}
	Following \cite{rohdin24_asvspoof}, the ASVspoof 5 dataset was used for training. A ResNet34 generated ASV log-likelihood ratios (LLRs) for speaker verification, while the {\it BWRFN-ResNet} model produced countermeasure (CM) LLRs for anti-spoofing. The SASV system fuses both LLRs to compute the final SASV LLR, using the fusion and calibration process from \cite{rohdin24_asvspoof}.
	
	\section{Experiments and Analysis}
	This section presents the experimental results and analysis of domain generalization performance. For cross-dataset ASV evaluation, the EER is reported after the final training epoch. For the closed condition of Track~1, both minDCF and EER were used as evaluation metrics, while the closed condition of Track~2 was assessed using min a-DCF, min t-DCF, and t-EER.
	
	\begin{table}[!h]
		\caption{EERs (\%) of the proposed method and other baselines on the cross-dataset ASV task. WRFN-R-vector represents weighted RFN-R-vector, where the weights were optimized during training. A comprehensive evaluation is conducted by combining both seen and unseen trials into a unified set of assessment trials, referred to as ``Overall''.}
		\centering
		\label{tab4}
		\setlength{\tabcolsep}{2pt} 
		\begin{tabular}{lcccc}
			\hline
			\multirow{2}{*}{Methods} & \multicolumn{2}{c}{Seen} & Unseen & \multirow{2}{*}{Overall} \\
			& FFSVC     & CN-Celeb     & HI-MIA &                          \\
			\cmidrule(lr){1-1} \cmidrule(lr){2-3} \cmidrule(lr){4-4} \cmidrule(lr){5-5} 
			R-vector                 & 4.56      & 16.01        & 12.65  & 13.36                    \\
			ECAPA-TDNN               & 9.67      & 23.04        & 12.30  & 16.84                    \\
			RFN-R-vector \cite{kim2022domain}     & 4.39      & 16.30        & 12.96  & 13.64                    \\
			WRFN-R-vector            & 4.26      & \bf 16.08        & 14.16  & 13.05                    \\
			\bf BWRFN-R-vector(ours)      & \bf 4.24      & 16.96        & \bf 8.15   & \bf 12.38                    \\ \hline
		\end{tabular}
	\end{table}
	Table \ref{tab4} presents the performance evaluations across various datasets. Multiple datasets with diverse domain shifts are utilized to better simulate real-world complexities. The results show that the BWRFN-R-vector achieves the lowest EER in the seen FFSVC and unseen HI-MIA. This improvement stems from the BWRFN-R-vector’s ability to model more intricate relationships between different normalizers. These findings highlight that the Bayesian-based weight modeling in BWRFN-R-vector enhances both the robustness and accuracy of speaker verification, particularly for out-of-distribution datasets like HI-MIA. Additionally, our method outperforms WRFN-R-vector, especially in unseen domains. This demonstrates that Bayesian learning methods surpass learnable weighted methods in cross-dataset evaluation.
	
	\begin{table}[!h]
		\caption{EERs (\%) of our proposed method and state-of-the-art methods on the
			cross-dataset ASV task. We inserted the BWRFN into different residual blocks (L1---L4) of the R-vector.}
		\centering
		\label{tab5}
		\begin{tabular}{lc}
			\hline
			Method         & Overall \\ \hline
			R-vector       & 13.36   \\
			R-vector+BWRFN-L1(ours) & 12.83   \\
			R-vector+BWRFN-L2(ours) & \bf 12.38   \\
			R-vector+BWRFN-L3(ours) & 12.58   \\
			R-vector+BWRFN-L4(ours) & 12.96   \\ \hline
		\end{tabular}
	\end{table}
	We conduct experiments to find which layer is better for inserting the BWRFN operations, and the results are reported in Table \ref{tab5}. The varying performance across the residual blocks suggests that the position of BWRFN integration within the R-vector framework can influence the overall effectiveness, with L2 being the optimal residual block for this particular task.
	
	\begin{table}[!h]
		
		\caption{Results of BWRFN-Resnet on the development set of the closed condition of Track 1 of ASVspoof 5.}
		\centering
		\label{tab6}
		\begin{tabular}{lcc}
			\hline
			Methods        & minDCF  & EER    \\ \hline
			SincNet-ASSIST \cite{falez24_asvspoof} & 0.35   & 13.71  \\
			ResNet-18 \cite{dao24_asvspoof} & 0.20 & 12.15 \\ 
			ResNet-34 \cite{dao24_asvspoof}     & 0.18    & 11.84  \\
			binspf \cite{rohdin24_asvspoof}         & 0.14  & 12.16 \\
			\bf BWRFN-ResNet (ours)          & \bf 0.13 & \bf 11.36 \\ \hline
		\end{tabular}
	\end{table}
	Table \ref{tab6} shows that the performance of our proposed method achieves superior performance on anti-spoofing compared with other state-of-the-art methods. This improvement highlights the effectiveness of our method in addressing domain mismatch for different unknown TTS algorithms. The good performance is attributed to the ability of Bayesian learning, which can account for the parameter uncertainty and generalize better for the unseen domain.
	
	\begin{table}[!h]
		
		\caption{Results of BWRFN-Resnet on the development set of the closed condition of Track~2 of ASVspoof 5.}
		\centering
		\label{tab:tab7}
		\setlength{\tabcolsep}{3pt} 
		\begin{tabular}{lccc}
			\hline
			Methods           & min a-DCF                    & min t-DCF                    & t-EER                       \\ \hline
			ECAPA-TDNN+ASSIST \cite{kurnaz24_asvspoof} & 0.1692                       &  N/A                            & N/A                            \\
			AASIST3 \cite{borodin24_asvspoof}           &   N/A                           & 0.2657                       &  N/A                           \\
			FwSE-ResNet100 \cite{villalba24_asvspoof}    & 0.134                        & 0.219                        & 6.53                        \\
			CM\#5,ASV\#2 \cite{rohdin24_asvspoof}     & 0.127 & 0.209 & 6.57 \\
			\bf BWRFN-ResNet (ours)              & \bf 0.125 & \bf 0.205 & \bf 5.83 \\ \hline
		\end{tabular}
	\end{table}
	Our method outperforms the current state-of-the-art approaches for the closed condition of Track 2 of ASVspoof 5, as shown in Table~\ref{tab:tab7}. This demonstrates the effectiveness of our approach in tackling the cross-domain problem and highlights its advantages in domain generalization. Compared with other methods, our model achieves significant improvements, reflecting its ability to better capture domain-invariant features. 
	
	\section{Conclusions and Future Work}
	In this paper, we have introduced the BWRFN-R-vector network, which integrates Bayesian learning into the weighting of the frequency components in the relaxed instance frequency-wise normalization. Results demonstrate that accounting for the uncertainty in the frequency-dependent weights can mitigate domain generalization issues. In future work, we plan to extend this approach to other front-end models.
	
	\section{Acknowledgements}
	This work was supported by the RGC of Hong Kong SAR, Grant No. PolyU 15228223 and the PolyU's Research Student Attachment Programme. The work was supported by Czech Ministry of Education, Youth and Sports (MoE) through the OP JAK project ``Linguistics, Artificial Intelligence and Language and Speech Technologies: from Research to Applications'' (ID:CZ.02.01.01/00/23\_020/0008518). Computing on IT4I supercomputer was supported by MoE through the e-INFRA CZ (ID:90254).

	\bibliographystyle{IEEEtran}
	\bibliography{mybib}
	
\end{document}